\documentclass[12pt]{article}
\usepackage[dvips]{graphicx}
\def\beq{\begin{equation}}
\def\eeq{\end{equation}}
\def\bea{\begin{eqnarray}}
\def\eea{\end{eqnarray}}
\def\ben{\begin{enumerate}}
\def\een{\end{enumerate}}

\def\bar{\overline}

\def\a{\alpha}

\def\l{\lambda}
\def\bc{\begin{center}}
\def\ec{\end{center}}
\def\O{{\cal O}}

\renewcommand{\theequation}{\arabic{equation}}

\def\PRD#1#2#3{Phys. Rev. D {\bf #1} (#3), #2}

\def\PLB#1#2#3{Phys. Lett. B {\bf #1} (#3), #2}
\def\NPB#1#2#3{Nucl. Phys. B {\bf #1} (#3), #2}

\def\PTP#1#2#3{Prog. Theor. Phys. {\bf #1} (#3), #2}

\def\JHEP#1#2#3{J. High Energy Phys. {\bf #1} (#3), #2}
\begin{document} 

\begin{flushright}
hep-ph/0309302 \qquad AUE-03-01 / KGKU-03-01 
\end{flushright}

\vspace{3mm}

\begin{center}
{\large \bf Particle Spectra and Gauge Unification \\
          in $SU(6) \times SU(2)_R$ Model }

\vspace{10mm}

Takemi HAYASHI,$^{1, }$
            \footnote{E-mail: hayashi@kogakkan-u.ac.jp} 
Masahisa MATSUDA$^{2, }$
            \footnote{E-mail: mmatsuda@auecc.aichi-edu.ac.jp} 
and Takeo MATSUOKA$^{3, }$
            \footnote{E-mail: matsuoka@kogakkan-u.ac.jp} 

\end{center}

\vspace{10mm}

\begin{center}
{\it 
{}$^1$Kogakkan University, Ise 516-8555, Japan \\
{}$^2$Department of Physics and Astronomy, Aichi University 
of Education, Kariya 448-8542, Japan \\
{}$^3$Kogakkan University, Nabari 518-0498, Japan 
}
\end{center}

\vspace{3mm}

\begin{abstract}
We study the path from the string scale physics 
to the low-energy physics in the $SU(6) \times SU(2)_R$ 
string-inspired model with the flavor symmetry 
${\bf Z}_M \times {\bf Z}_N \times \tilde{D}_4$. 
The flavor symmetry controls the mass spectra of 
heavy particles as well as those of quarks and leptons 
in the intermediate energy region ranging from the string scale 
($\sim 10^{18} \, {\rm GeV}$) to the electroweak scale. 
In this paper we examine the mass spectra of heavy particles 
in detail in our model. 
The renormalization group evolution of the gauge couplings 
is studied up to two-loop order. 
A consistent solution of the gauge unification around 
the string scale is found by adjusting the spectra 
of the anti-generation matter fields. 
\end{abstract}

\newpage 
\section{Introduction}

The study of the path for connecting the string theory 
with the low-energy physics is one of the most important issues 
in particle physics and cosmology. 
However, we are still in the early stages of the study. 
In fact, we do not yet have a comprehensive string-based 
understanding of apparent characteristic patterns in quark 
and lepton masses and mixings at low energies. 
It is considered that these characteristic patterns are closely 
linked to the flavor symmetry, which is expected to arise 
from the symmetric structure of the compact space 
in the string theory. 
It is also likely that the flavor symmetry controls 
the mass spectra of heavy particles in the intermediate 
energy region as well as those of quarks and leptons.

Recent developments in the string theory have provided 
new aspects of string phenomenology. 
It has been pointed out in Ref.\cite{Fuzz} that 
a new type of non-Abelian flavor symmetry can appear additionally 
if the compact space is non-commutative. 
As a matter of fact, in a string with discrete torsion, 
the coordinates become non-commutative 
operators.\cite{Torsion,Torsion2,Beren} 
As a phenomenological approach, in Refs.\cite{Fuzz} and \cite{Anom} 
the flavor symmetry ${\bf Z}_M \times {\bf Z}_N \times \tilde{D}_4$ 
has been introduced into the string-inspired model, 
where the cyclic group ${\bf Z}_M$ and the binary dihedral group 
$\tilde{D}_4$ have R symmetries, while ${\bf Z}_N$ 
has not.\footnote{
We use here the notation "$\sim$" for the binary group. 
In Ref.\cite{Anom} we have discussed the projective representation 
of the dihedral group $D_4$, which is identical with the binary 
dihedral group $\tilde{D}_4$. 
} 
The introduction of a binary representation of non-Abelian flavor 
symmetry is motivated also by the phenomenological observation that 
the R-handed Majorana neutrino mass for the third generation 
is nearly equal to the geometrical average of the string scale 
$M_S$ ($\sim 10^{18} \, $GeV) and the electroweak scale $M_Z$. 
In Ref.\cite{Anom}, solving the anomaly-free conditions under 
many phenomenological constraints coming from the particle spectra, 
we were led to a large mixing angle (LMA)-MSW solution with 
$(M, \ N)=(19, \ 18)$, in which the appropriate flavor charges 
are assigned to the matter fields. 
The solution includes phenomenologically acceptable results 
concerning fermion masses and mixings and also concerning 
hierarchical energy scales including the string scale, $\mu$ scale 
and the Majorana mass scale of R-handed neutrinos.

In the framework of the string theory, we are prohibited from 
adding extra matter fields by hand. 
This situation is in sharp contrast to that of 
the conventional GUT-type models. 
In fact, in the perturbative heterotic string theory 
we have no adjoint or higher representation matter (Higgs) fields. 
Within this rigid framework, we have discussed the path 
from the string scale physics to the low-energy physics 
in the $SU(6) \times SU(2)_R$ string-inspired model 
with the flavor symmetry 
${\bf Z}_{19} \times {\bf Z}_{18} \times \tilde{D}_4$. 
In Ref.\cite{Preco} the renormalization group (RG) equations down 
from the $M_S$ have been studied for the scalar masses squared 
of the gauge non-singlet matter fields. 
It has been found that the radiative breaking of the gauge 
symmetry can occur slightly below the $M_S$ 
due to the large Yukawa coupling 
which is identical with the colored Higgs coupling. 
This symmetry breaking triggers a subsequent symmetry 
breaking.\cite{Scale} 
Then, we obtain the sequential symmetry breaking 
\[
  SU(6) \times SU(2)_R \longrightarrow 
   SU(4)_{\rm PS} \times SU(2)_L \times SU(2)_R \longrightarrow 
   G_{\rm SM},
\]
where $SU(4)_{\rm PS}$ and $G_{\rm SM}$ represent 
the Pati-Salam $SU(4)$\cite{Pati} and the standard model 
gauge group, respectively.

The purpose of this paper is to pursue the further 
exploration of the path from the string scale physics 
to the low-energy physics. 
For this purpose we examine the particle spectra 
in the intermediate energy region in detail 
in the $SU(6) \times SU(2)_R$ model with the flavor symmetry 
${\bf Z}_{19} \times {\bf Z}_{18} \times \tilde{D}_4$. 
Afterward, the RG runnings of the gauge couplings are 
studied up to two-loop order. 
It should be emphasized that in our model 
the unification scale is not the so-called GUT scale 
($\sim 2 \times 10^{16} \, {\rm GeV})$ but around the string scale 
($\sim 10^{18} \, {\rm GeV})$. 
We find a solution of the gauge coupling unification 
by adjusting the spectra of the anti-generation matter 
fields, with which the K\"ahler class moduli fields couple. 
In this solution $SU(3)_c$ and $SU(2)_L$ gauge couplings meet 
at $\O(5 \times 10^{17} \, {\rm GeV})$. 
However, $SU(6)$ and $SU(2)_R$ gauge couplings are not 
perturbatively unified at the string scale. 
We expect that the non-perturbative unification of 
$SU(6)$ and $SU(2)_R$ gauge couplings is properly realized 
in the framework of the higher-dimensional underlying theory.

This paper is organized as follows. 
In \S 2, we explain main features of the $SU(6) \times SU(2)_R$ 
model with the flavor symmetry 
${\bf Z}_{19} \times {\bf Z}_{18} \times \tilde{D}_4$. 
Although we have never yet found a concrete example of the string 
compactification which induces this type of the flavor symmetry 
exactly, 
${\bf Z}_M \times {\bf Z}_N$ type of the flavor 
symmetry appears in some kinds of the Calabi-Yau 
compactification.\cite{4gen,Gepner} 
Further, the binary type of the flavor symmetry is expected 
to stem from the compact space with non-commutative geometry. 
In this paper we assume 
${\bf Z}_{19} \times {\bf Z}_{18} \times \tilde{D}_4$ 
as the flavor symmetry. 
In \S 3, we examine the particle spectra in the intermediate 
energy region ranging from the $M_S$ to the $M_Z$. 
In the intermediate energy region there appear rich spectra 
of extra heavy particles beyond the minimal supersymmetric 
standard model.
After presenting the two-loop RG equations of the gauge 
couplings in the intermediate region, 
we carry out the numerical analysis of the RG runnings 
of the gauge couplings in \S 4. 
We explore solutions of the gauge coupling unification. 
The final section is devoted to summary and discussion. 
In Appendix we explain dominant effective Yukawa couplings 
contributing to the RG evolution of the gauge couplings 
in the intermediate energy region. 

\vspace{10mm}

\section{$SU(6) \times SU(2)_R$ model with the flavor symmetry}

Let us start by briefly summarizing the main points of 
the $SU(6) \times SU(2)_R$ string-inspired model considered here. 
More detailed descriptions are given in 
Refs.\cite{Fuzz,Anom,Preco,Matsu1,Matsu2,Matsu3,CKM,MNS}. 

\begin{enumerate}
\item The gauge group is assumed to be $G = SU(6) \times SU(2)_R$, 
which can be obtained from $E_6$ through the ${\bf Z}_2$ 
flux breaking on a multiply-connected manifold 
$K$.\cite{Hoso,Flux1,Flux2} 
In contrast to the conventional GUT-type models, 
we have no Higgs fields of adjoint or higher representations. 
Nevertheless, the symmetry breaking of $G$ down to $G_{\rm SM}$ 
can take place via the Higgs mechanism.\cite{Matsu4} 

\item In the context of the string theory, it is assumed 
that the gauge non-singlet matter content consists of the chiral 
superfields of three families and the single vector-like multiplet 
in the form 
\begin{equation}
  3 \times {\bf 27}(\Phi_{1,2,3}) + 
        ({\bf 27}(\Phi_0)+\overline{\bf 27}({\bar \Phi})) 
\end{equation}
in terms of $E_6$. 
The superfields $\Phi$ in {\bf 27} of $E_6$ are decomposed into 
two irreducible representations of $G = SU(6) \times SU(2)_R$ as 
\begin{equation}
  \Phi({\bf 27})=\left\{
       \begin{array}{lll}
         \phi({\bf 15},{\bf 1})& : 
               & \quad \mbox{$Q,L,g,g^c,S$}, \\
          \psi({\bf 6}^*,{\bf 2}) & : 
               & \quad \mbox{$(U^c,D^c),(N^c,E^c),(H_u,H_d)$}, 
       \end{array}
       \right.
\label{eqn:27}
\end{equation}
where the pair $g$ and $g^c$ and the pair $H_u$ and $H_d$ represent 
the colored Higgs and the doublet Higgs superfields, respectively, 
$N^c$ is the R-handed neutrino superfield, and 
$S$ is an $SO(10)$ singlet. 
It should be noted that the doublet Higgs and the color-triplet 
Higgs fields belong to different irreducible representations 
of $G$, as shown in Eq. (\ref{eqn:27}). 
As a consequence, the triplet-doublet splitting problem is 
solved naturally.\cite{Matsu1} 
In our model there are only two types of gauge invariant trilinear 
combinations 
\bea
    (\phi ({\bf 15},{\bf 1}))^3 & = & QQg + Qg^cL + g^cgS, \\
    \phi ({\bf 15},{\bf 1})(\psi ({\bf 6}^*,{\bf 2}))^2 & 
            = & QH_dD^c + QH_uU^c + LH_dE^c  + LH_uN^c 
                                            \nonumber \\ 
             {}& & \qquad   + SH_uH_d + 
                     gN^cD^c + gE^cU^c + g^cU^cD^c. 
\label{eqn:trico}
\eea

\item As the flavor symmetry, we introduce 
the ${\bf Z}_{19} \times {\bf Z}_{18}$ and $\tilde{D}_4$ symmetries 
and regard ${\bf Z}_{19}$ and ${\bf Z}_{18}$ as the R and non-R 
symmetries, respectively. 
$\tilde{D}_4$ represents the binary dihedral group. 
Because the numbers 19 and 18 are relatively prime, 
we can combine these symmetries as 
\begin{equation}
  {\bf Z}_{19} \times {\bf Z}_{18} = {\bf Z}_{342}. 
\end{equation}
Solving the anomaly-free conditions under the many 
phenomenological constraints coming from the particle spectra, 
we obtain a LMA-MSW solution with the ${\bf Z}_{342}$ charges 
of the matter superfields, as shown in Table 1.\cite{Anom} 
In this solution we assign the charge $(-1, \ 0)$ under 
${\bf Z}_{19} \times {\bf Z}_{18}$, {\it i.e.}, the charge 
18 under ${\bf Z}_{342}$ to the Grassmann number $\theta$. 
The assignment of the ``$\tilde{D}_4$ charges" to the matter 
superfields is given in Table 2, 
where $\sigma_i \, (i=1,2,3)$ represent the Pauli matrices and 
\begin{equation}
   \sigma_4 =  \left(
       \begin{array}{cc}
         1  &  0  \\
         0  &  i  
       \end{array}
       \right). 
\end{equation} 
The $\sigma_3$ transformation yields the R-parity. 
It is found that the R-parities of the superfields $\Phi_i \, 
(i=1,2,3)$ for the three generations are all odd, 
while those of $\Phi_0$ and $\bar{\Phi}$ are even. 

\end{enumerate}

\vspace{5mm}

\begin{table}[t]
\caption{Assignment of ${\bf Z}_{342}$ charges for matter superfields. 
$\theta$ has the charge 18.}
\label{table:1}
\begin{center}
\begin{tabular}{|c|ccc|cc|} \hline \hline 
  & \phantom{M} $\Phi_1$ \phantom{MM} & 
      \phantom{MM} $\Phi_2$ \phantom{MM} & 
        \phantom{MM} $\Phi_3$ \phantom{M} & 
          \phantom{M} $\Phi_0$ \phantom{MM} & 
            \phantom{MM} $\bar{\Phi}$ \phantom{M} \\ \hline
$\phi({\bf 15, \ 1})$   &  $a_1=126$  &  $a_2=102$  &  $a_3=46$  
                                &  $a_0=12$  &  $\bar{a}=-16$  \\
$\psi({\bf 6^*, \ 2})$  &  $b_1=120$  &  $b_2=80$  &  $b_3=16$  
                         &  $b_0=-14$  &  $\bar{b}=-67$  \\ \hline
\end{tabular}
\end{center}
\end{table}

\begin{table}[t]
\caption{Assignment of ``$\tilde{D}_4$ charges" to matter superfields. 
The ``$\tilde{D}_4$ charge" of $\theta$ is $\sigma_1$.}
\label{table:2}
\begin{center}
\begin{tabular}{|c|ccc|} \hline \hline 
    & \phantom{M} $\Phi_i \ (i=1,2,3)\ $ \phantom{M} & 
        \phantom{M} $\Phi_0$ \phantom{M} & 
              \phantom{M} $\bar{\Phi}$ \phantom{M} \\ \hline
$\phi({\bf 15, \ 1})$   & 
                   $\sigma_1$  &     1       &       1       \\
$\psi({\bf 6^*, \ 2})$  & 
                   $\sigma_2$  & $\sigma_3$  &   $\sigma_4$  \\  \hline
\end{tabular}
\end{center}
\end{table}

Due to the gauge symmetry and the flavor symmetry, 
the superpotential terms which induce the effective Yukawa 
couplings at low energies take the forms 
\bea
  W_Y & = & \frac{1}{3!} \, z_0 \left( \frac{\phi_0 \bar{\phi}}
              {M_1^2} \right)^{\zeta_0} (\phi_0)^3 
          + \frac{1}{3!} \, \bar{z} \left( \frac{\phi_0 \bar{\phi}}
              {M_1^2} \right)^{\bar{\zeta}} (\bar{\phi})^3 
          + \frac{1}{2} \, h_0 \left( \frac{\phi_0 \bar{\phi}}{M_1^2} 
               \right)^{\eta_0} \phi_0 \psi_0 \psi_0   \nonumber \\
     {} & & \phantom{M} + \frac{1}{2} \, \bar{h} \left( 
               \frac{\phi_0 \bar{\phi}}{M_1^2} \right)^{\bar{\eta}} 
                  \left( \frac{\psi_0 \bar{\psi}}{M_2^2} \right)^2 
                    \bar{\phi} \, \bar{\psi} \, \bar{\psi} 
           + \frac{1}{2} \, \sum_{i,j=1}^{3} z_{ij} 
                  \left( \frac{\phi_0 \bar{\phi}}{M_1^2} 
                  \right)^{\zeta_{ij}} \phi_0 \phi_i \phi_j   \nonumber \\
     {}& & \phantom{M} + \frac{1}{2} \, \sum_{i,j=1}^{3} h_{ij} \left( 
           \frac{\phi_0 \bar{\phi}}{M_1^2} \right)^{\eta_{ij}} 
                          \phi_0 \psi_i \psi_j 
     + \sum_{i,j=1}^{3} m_{ij} \left( \frac{\phi_0 \bar{\phi}}
              {M_1^2} \right)^{\mu_{ij}} \psi_0 \phi_i \psi_j, 
\label{eqn:WY}
\eea
where $\phi_0 \bar{\phi}$ and $\psi_0 \bar{\psi}$ stand for 
the gauge-singlet combinations. 
The scale $M_1$ ($M_2$) represents the string scale $M_S$ multiplied 
by the $\O(1)$ factor coming from the volume of the compact space 
in which the matter fields $\phi_0({\bf 15, \ 1})$ and 
$\bar{\phi}({\bf 15^*, \ 1})$ ($\psi_0({\bf 6^*, \ 2})$ and 
$\bar{\psi}({\bf 6, \ 2})$) reside. 
The exponents are determined by the constraints coming from 
the flavor symmetry and concretely given by 
\bea
    (\zeta_0, \ \bar{\zeta}, \ \eta_0, \ \bar{\eta}) 
              & = & (0, \ 150, \ 158, \ 84), \qquad 
    \zeta_{ij} =  \left(
       \begin{array}{ccc}
         57  &  51  &  37  \\
         51  &  45  &  31  \\
         37  &  31  &  17  
       \end{array}
       \right)_{ij},       \nonumber \\
    \eta_{ij} & = & \left(
       \begin{array}{ccc}
         54  &  44  &  28  \\
         44  &  34  &  18  \\
         28  &  18  &   2  
       \end{array}
       \right)_{ij},        \quad \ 
    \mu_{ij} =  \left(
       \begin{array}{ccc}
         49  &  39  &  23  \\
         43  &  33  &  17  \\
         29  &  19  &   3  
       \end{array}
       \right)_{ij}. 
\label{eqn:power}
\eea
The coefficients $z_0$, $\bar{z}$, $h_0$, $\bar{h}$, $z_{ij}$, 
$h_{ij}$ and $m_{ij}$ are $\O(1)$ constants. 
The relation $\zeta_0 = 0$ means that only the superfield 
$\phi_0$ takes part in the renormalizable interaction with the 
large Yukawa coupling at $M_S$. 
All or some of the powers of the gauge-singlet combination 
$\phi_0 \bar{\phi}$ can be replaced with those of another gauge-singlet 
combination $\psi_0 \bar{\psi}$ subject to the flavor symmetry.

Here it is important for us to comment on the role of 
the moduli fields. 
In the Calabi-Yau string, the generation matter and 
the anti-generation matter couple separately with 
the complex structure moduli fields and 
the K\"ahler class moduli fields, respectively, 
in the superpotential.\cite{Dixon} 
The nonvanishing vacuum expectation values (VEVs) of 
the moduli fields are expected to be $\O(M_S)$. 
The VEVs of the K\"ahler class moduli fields represent 
the size and shapes of the compact space. 
In addition, the K\"ahler class moduli fields carry the flavor 
charges and their VEVs control the flavor symmetry. 
Therefore, there is a possibility that the second and the fourth 
terms in Eq. (\ref{eqn:WY}) are supplemented with the other terms 
multiplied by a certain function of the K\"ahler class moduli fields. 
In the next section we discuss the possible modification of 
the second term.

When $\phi_0$ and $\bar{\phi}$ develop non-zero VEVs, 
the above non-renormalizable terms induce effective Yukawa 
couplings with hierarchical patterns. 
Namely, below the scale $|\langle \phi_0 \rangle|$, 
the Froggatt-Nielsen mechanism acts for non-renormalizable 
interactions.\cite{F-N} 
in the superpotential 
Further, we have another non-renormalizable terms 
\begin{equation}
  W_1 = M_1^3 \left[ \l_0 
     \left( \frac{\phi_0 \bar{\phi}}{M_1^2} \right)^{2n} 
       + \l_1 \left( \frac{\phi_0 \bar{\phi}}{M_1^2} \right)^n 
         \left( \frac{\psi_0 \bar{\psi}}{M_2^2} \right)^m 
       + \l_2 \left( \frac{\psi_0 \bar{\psi}}{M_2^2} \right)^{2m}\right] 
\label{eqn:W1}
\end{equation}
with $\l_i = \O(1)$. 
The flavor symmetry ${\bf Z}_{342} \times \tilde{D}_4$ requires 
$n = 81$ and $m=4$. 
We assume that the supersymmetry is broken at the string 
scale due to the hidden sector dynamics and that 
the supersymmetry (SUSY) breaking is communicated gravitationally 
to the observable sector via the universal soft SUSY breaking terms. 
The scale of the SUSY breaking $\tilde{m}_{\phi}$ is supposed 
to be $10^3 \,{\rm GeV}$. 
Under this assumption we study the minimum point of 
the scalar potential.\cite{Scale} 
When we have a large coupling for the Yukawa interaction 
$z_0 \, \phi_0^3$ at the string scale $M_S$, 
through the RG evolution the scalar masses squared become 
negative slightly below $M_S$.\cite{Preco} 
As a consequence, the ground state is characterized by 
$\langle \phi_i \rangle = \langle \psi_i \rangle = 0, \ (i=1,2,3)$ 
and 
\beq
   \frac{|\langle \phi_0 \rangle|}{M_1} = 
     \frac{|\langle \bar{\phi} \rangle|}{M_1} = 0.894, \qquad 
   \frac{|\langle \psi_0 \rangle|}{M_2} = 
     \frac{|\langle \bar{\psi} \rangle|}{M_2} = 0.103, 
\eeq
where we take the numerical values 
$M_1 = M_2 = 5 \times 10^{17} \,{\rm GeV} = M_S/3$ and 
\beq
   \left( \frac{|\langle \phi_0 \rangle|}{M_1} \right)^{4n-2}
     \, \times M_1 = c \, \tilde{m}_{\phi}.
\eeq
The coefficient $c$ in the r. h. s. is expressed as 
a function of $\l_{0,1,2}$, $n$ and $m$ and 
we take here $c=0.1$. 
In this vacuum we have the relations 
\beq
    x^{81} \simeq y^4, \qquad x^{161} \, M_1 = 10^2 \, {\rm GeV}, 
\eeq
where we use the notation 
$x = (|\langle \phi_0 \rangle|/M_1)^2$ and 
$y = (|\langle \psi_0 \rangle|/M_2)^2$. 
Thus, below the scale $|\langle \psi_0 \rangle|$ we obtain the 
effective Yukawa superpotential 
\bea
  W^{(eff)}_Y & = & \frac{1}{3!} \, z_0 \, x^{\zeta_0} (\phi_0)^3 
          + \frac{1}{3!} \, \bar{z} \, x^{\bar{\zeta}} (\bar{\phi})^3 
          + \frac{1}{2} \, h_0 \, x^{\eta_0} \phi_0 \psi_0 \psi_0   \nonumber \\
     {} & & \phantom{M} + \frac{1}{2} \, \bar{h} \, x^{\bar{\eta}} 
                 \, y^2 \bar{\phi} \, \bar{\psi} \, \bar{\psi} 
           + \frac{1}{2} \, \sum_{i,j=1}^{3} z_{ij} \, 
                  x^{\zeta_{ij}} \phi_0 \phi_i \phi_j   \nonumber \\
     {}& & \phantom{M} + \frac{1}{2} \, \sum_{i,j=1}^{3} h_{ij} \, 
                  x^{\eta_{ij}} \phi_0 \psi_i \psi_j 
     + \sum_{i,j=1}^{3} m_{ij} \, x^{\mu_{ij}} \psi_0 \phi_i \psi_j. 
\label{eqn:WYeff}
\eea

\vspace{10mm}

\section{Mass spectra of heavy particles}
In this section we explore the particle spectra in the intermediate 
energy region ranging from the $M_S$ to the $M_Z$ 
for the R-parity even superfields first and then 
for the odd superfields. 

\vspace{10mm}

\noindent
A. \ The R-parity even superfields 

\vspace{5mm}

The R-parity even superfields contain $\phi_0$, ${\bar \phi}$, 
$\psi_0$ and ${\bar \psi}$. 
As mentioned above, the gauge symmetry is spontaneously broken 
at the scale 
$|\langle \phi_0({\bf 15, \ 1})\rangle | 
         \simeq 4.5 \times 10^{17} \, {\rm GeV}$, 
and subsequently at the scale 
$|\langle \psi_0({\bf 6^*, \ 2})\rangle | 
         \simeq 5.2 \times 10^{16} \, {\rm GeV}$. 
This yields the symmetry breakings 
\begin{equation}
   SU(6) \times SU(2)_R 
     \buildrel \langle \phi_0 \rangle \over \longrightarrow 
             SU(4)_{\rm PS} \times SU(2)_L \times SU(2)_R  
             \buildrel \langle \psi_0 \rangle \over \longrightarrow 
     G_{\rm SM}. 
\end{equation}
Since the fields that develop non-zero VEVs are singlets under 
the remaining gauge symmetries, 
they are assigned as 
$\langle \phi_0({\bf 15, \ 1})\rangle = \langle S_0 \rangle $ and 
$\langle \psi_0({\bf 6^*, \ 2})\rangle = \langle N^c_0 \rangle $. 
In the first step of the symmetry breaking, the fields $Q_0$, $L_0$, 
${\overline Q}$, ${\overline L}$ and $(S_0 - {\overline S})/\sqrt{2}$ 
are absorbed by the gauge fields. 
Through the subsequent symmetry breaking, 
the fields $U_0^c$, $E_0^c$, ${\overline U}^c$, 
${\overline E}^c$ and $(N_0^c - {\overline N}^c)/\sqrt{2}$ 
are absorbed. 
Therefore, below the scale $|\langle \psi_0 \rangle|$ 
the remaining modes in the R-parity even superfields are 
gauge singlet fields ($S' \equiv (S_0 + {\overline S})/\sqrt{2}$, 
$N'^c \equiv (N_0^c + {\overline N}^c)/\sqrt{2}$), 
doublet Higgs fields ($H_{u_0}$, $H_{d_0}$, ${\bar H_u}$, ${\bar H_d}$) 
and down-type colored fields ($g_0$, $g_0^c$, $D_0^c$, 
${\overline g}$, ${\overline g^c}$, ${\overline D^c}$). 

\vspace{5mm}

\ben
\item Gauge singlet fields $S'$ and $N'^c$ \\
Mass matrix for $S'$ and $N'^c$ is induced from 
Eq. (\ref{eqn:W1}) as 
\beq
  \begin{array}{r@{}l} 
    \vphantom{\bigg(}   &  \begin{array}{ccc} 
      & \phantom{MMMM}   S'   &  \phantom{MMMMMMMM}  N'^c  
                           \end{array}  \\ 
\widehat{{\cal M}}_{S' N'} 
    = 
   \begin{array}{l} 
        S'  \\
        N'^c  \\ 
   \end{array} 
   & 
\left( 
    \begin{array}{cc} 
      \l_0 \O(4n^2) x^{2n-1}    &   \l_1 \O(nm) x^{2n(1-1/4m)-1/2}       \\
     \l_1 \O(nm) x^{2n(1-1/4m)-1/2}   &  \l_2 \O(4m^2)) x^{2n(1-1/2m)} 
    \end{array} 
      \right),  
  \end{array} 
\label{eqn:SN}
\eeq
in $M_1$ units, where $n=81$ and $m=4$. 
The eigenvalues are given by 
\beq
  \O(x^{118} \, M_1) \sim 10^{6.2} \, {\rm GeV}, \qquad 
  \O(x^{125} \, M_1) \sim 10^{5.5} \, {\rm GeV}. 
\eeq
In the following we use the notation $\tilde{S}$ and $\tilde{N^c}$ 
for the mass eigenstates. 

\vspace{5mm}

\item Doublet Higgs fields $H_{u_0}$, $H_{d_0}$, 
${\overline H_u}$ and ${\overline H_d}$ \\ 
The mass matrix for these fields is derived from the terms 
in Eqs. (\ref{eqn:W1}) and (\ref{eqn:WYeff}). 
The effective superpotential contributing to this mass matrix 
becomes 
\beq
 W_H^{(eff)}  \simeq  h_0 \, x^{\eta_0} \, S_0 H_{u0} H_{d0} 
        + {\bar h} \, x^{\bar \eta} \, y^2 \, \bar{S}\bar{H_u}\bar{H_d}  
       + \l_1 \, x^n \, y^{m-1} \, (H_{u_0}\bar{H_u}+H_{d_0}\bar{H_d}),
\eeq
where the first term induces the so-called $\mu$-term with 
$\mu = \O( x^{\eta_0 + 1/2} \, M_1)$. 
Using $(\eta_0, \ \bar{\eta}) = (158, \ 84)$ 
and $x^{81} \simeq y^4$, we obtain the mass matrix 
\beq
\begin{array}{r@{}l} 
   \vphantom{\bigg(}   &  \begin{array}{ccc} 
          \qquad   H_{d_0}   &  \phantom{MM} \bar{H_u}  &  
                          \end{array}  \\ 
\widehat{{\cal M}}_{H} \simeq 
   \begin{array}{l} 
        H_{u_0}    \\
        \bar{H_d}  \\ 
   \end{array} 
     & 
\left( 
  \begin{array}{cc} 
      h_0 \, x^{158.5}     &   \l_1 \, x^{141.75}     \\
      \l_1 \, x^{141.75}   &   \bar{h} \, x^{125}
  \end{array} 
\right)
\end{array}
\label{eqn:Higgs}
\eeq
in $M_1$ units. 
The eigenvalues are 
\beq
  \O(x^{125} \, M_1)   \sim 10^{5.5} \, {\rm GeV}, \qquad 
  \O(x^{158.5} \, M_1) \sim 10^{2.2} \, {\rm GeV}. 
\eeq

\vspace{5mm}

\item Down-type colored fields $g_0$, $g_0^c$, $D_0^c$, 
${\overline g}$, ${\overline g^c}$ and ${\overline D^c}$ \\
The effective superpotential which yields the mass matrix 
for these fields is of the form 
\bea
  W_g^{(eff)} & \simeq &   z_0 \, x^{\zeta_0} \, S_0 g_0 g^c_0 
          + {\bar z} \, x^{\bar \zeta} \, \bar{S}\bar{g}\bar{g^c}
          + h_0 \, x^{\eta_0} \, g_0 N_0^c D_0^c 
          + {\bar h} \, x^{\bar \eta} \, \bar{g}\bar{N^c}\bar{D^c} \nonumber  \\
      & & + \l_0 \, x^{2n-1} \, (g_0\bar{g}+g_0^c\bar{g^c})
          + \l_1 \, x^n \, y^{m-1} \, D_0^c\bar{D^c}.
\label{eqn:Wg}
\eea
This effective superpotential leads to the mass matrix 
\beq
\begin{array}{r@{}l} 
   \vphantom{\bigg(}   &  \begin{array}{cccc} 
        \phantom{MM} g_0^c & \phantom{MM} \bar{g} & \phantom{MMM} D_0^c & 
                          \end{array}  \\ 
 \widehat{{\cal M}}_{g} \simeq 
   \begin{array}{l} 
        g_0  \\  \bar{g^c}  \\  \bar{D^c} \\
   \end{array} 
     & 
 \left( 
   \begin{array}{ccc} 
      z_0 \, x^{0.5}   &      \l_0 \, x^{161}     &  h_0 \, x^{168.125}    \\
      \l_0 \, x^{161}  &   {\bar z} \, x^{150.5}  &         0              \\
            0          &  {\bar h} \, x^{134.625} &  \l_1 \, x^{141.75}    \\
   \end{array} 
 \right)
\end{array}
\label{eqn:EggcDc}
\eeq
in $M_1$ units. 
Here, (2, 3) and (3, 1) elements in this matrix are 
not exactly zero but approximately zero. 
In fact, the terms $\bar{S} N^c_0 \bar{g^c} D^c_0$ 
and $S_0 \bar{N^c_0} g^c_0 \bar{D^c}$ are induced through 
the higher order effects and sufficiently suppressed 
compared to the terms in Eq. (\ref{eqn:Wg}). 
The eigenvalues of this matrix are $\O(x^{0.5} \, M_1)$, 
$\O(x^{134.625} \, M_1)$ and $\O(x^{157.625} \, M_1)$. 
So it turns out that one set of down-type colored superfields 
with even R-parity should exist around 
$x^{157.625} \, M_1 \simeq \O(10^{2.3} \, {\rm GeV})$. 
As pointed out in the previous section, however, 
in the Calabi-Yau string the anti-generation matter fields 
couple with the K\"ahler class moduli fields, 
which carries the flavor charge. 
Therefore, both the second and the fourth terms in 
Eq. (\ref{eqn:WYeff}), i.e., $\bar{\phi}^3$ and 
$\bar{\phi} \bar{\psi}^2$ terms, 
are possibly supplemented with the other terms. 
Since we have never yet known the definite content of the K\"ahler 
class moduli fields and also their flavor charges, 
the explicit form of the interactions of the K\"ahler class moduli 
fields is undetermined in the present approach. 
In this paper, from the phenomenological viewpoint, we consider 
a simple case that one of the K\"ahler class moduli 
fields couple with the anti-generation fields ${\bar \phi}$ 
only via the interaction 
\beq
   f(T) \left( \frac{S_0 \bar{S}}{M_1^2} \right)^{\bar{k}} 
                                               \bar{\phi}^3, 
\eeq 
where $T$ is one of the K\"ahler class moduli fields and carries 
an appropriate flavor charge. 
The exponent ${\bar{k}}$ is introduced as an unknown parameter, 
because we can not determine the flavor charge of $T$ at this stage 
as well as the explicit form of the function $f(T)$. 
In view of the fact that the K\"ahler class moduli fields 
represent the size and the shape of the compact space, 
it is supposed that the field $T$ develops a non-zero VEV 
with $\langle T \rangle=\O(M_S)$ and $f(\langle T \rangle)=\O(1)$. 
If ${\bar{k}} < 150.5$, 
a dominant term contributing to the (2, 2) element 
in Eq. (\ref{eqn:EggcDc}) is replaced by 
\beq
  f(\langle T \rangle) \, x^{\bar{k}} \, 
      \langle \bar{S} \rangle \bar{g}\bar{g^c} 
          = M_1 \, f(\langle T \rangle) \, x^k \, \bar{g}\bar{g^c}, 
\eeq
where $k = {\bar k}+0.5$. 
Then the mass matrix is also replaced by 
\beq
\begin{array}{r@{}l} 
   \vphantom{\bigg(}   &  \begin{array}{cccc} 
        \phantom{MM} g_0^c & \phantom{MM} \bar{g} & \phantom{MMM} D_0^c & 
                          \end{array}  \\ 
 \widehat{{\cal M}}_{g} \simeq 
   \begin{array}{l} 
        g_0  \\  \bar{g^c}  \\  \bar{D^c} \\
   \end{array} 
     & 
 \left( 
   \begin{array}{ccc} 
      z_0 \, x^{0.5}   &        \l_0 \, x^{161}       &  h_0 \, x^{168.125}    \\
      \l_0 \, x^{161}  & f(\langle T \rangle) \, x^k  &          0             \\
            0          &   {\bar h} \, x^{134.625}    &  \l_1 \, x^{141.75}    \\
   \end{array} 
 \right)
\end{array}
\label{eqn:EColor}
\eeq
in $M_1$ units. 
If we take $10.125 < k < 134.625$, 
the eigenvalues of this matrix become 
\beq
  \O(x^{0.5} \, M_1)    \sim 10^{17.7} \, {\rm GeV}, \quad 
  \O(x^k \, M_1),                                    \quad 
  \O(x^{141.75} \, M_1) \sim 10^{3.9} \, {\rm GeV} 
\eeq
with $x^{10.125} \, M_1 \sim 10^{16.7} \, {\rm GeV} > 
x^k \, M_1 > x^{134.625} \, M_1 \sim 10^{4.6} \, {\rm GeV}$. 
In the next section we adjust the parameter $k$ so as to 
find a consistent solution of gauge unification. 
Thus, the K\"ahler class moduli fields, although are gauge-singlets, 
possibly affect the evolution of the gauge couplings through 
the couplings with the anti-generation fields. 

\een

\vspace{5mm}

\begin{table}[t]
\caption{The multiplicities and the spectra of the R-parity even superfields. 
In this table the parameter $k$ is supposed to be $10.125 < k < 118$, 
because $k$ is adjusted as $k=42.5$ later.}
\label{table:3}
\begin{center}
\renewcommand{\arraystretch}{1.1}
\begin{tabular}{|ccc|ccccccc|} \hline \hline 
  & $\times M_1$  &   &   $Q$ & $L$ & $U^c$ & $E^c$ &  
   ($H_u$, $H_d$) & ($g$, $g^c$, $D^c$) & ($S$, $N^c$)   \\ \hline
 $1$          & $\sim$ & $x^{0.5}$     & 2 & 2 & 2 & 2 & 4 & 6 & 4   \\
 $x^{0.5}$    & $\sim$ & $x^{10.125}$  & 0 & 0 & 2 & 2 & 4 & 4 & 3   \\
 $x^{10.125}$ & $\sim$ & $x^k$         & 0 & 0 & 0 & 0 & 4 & 4 & 2   \\
 $x^k$        & $\sim$ & $x^{118}$     & 0 & 0 & 0 & 0 & 4 & 2 & 2   \\
 $x^{118}$    & $\sim$ & $x^{125}$     & 0 & 0 & 0 & 0 & 4 & 2 & 1   \\
 $x^{125}$    & $\sim$ & $x^{141.75}$  & 0 & 0 & 0 & 0 & 2 & 2 & 0   \\
 $x^{141.75}$ & $\sim$ & $x^{158.5}$   & 0 & 0 & 0 & 0 & 2 & 0 & 0   \\
 $x^{158.5}$  & $\sim$ &               & 0 & 0 & 0 & 0 & 0 & 0 & 0   \\
\hline
\end{tabular}
\end{center}
\vspace{3mm}
\end{table}

We summarize the multiplicities and the spectra of 
the R-parity even superfields in Table 3. 
We suppose here $10.125 < k < 118$, 
because the parameter $k$ is adjusted as $k=42.5$ later. 

\vspace{10mm}

\noindent
B. \ The R-parity odd superfields

\vspace{5mm}

The R-parity odd superfields contains three generations 
of matter superfields $\phi_i$ and $\psi_i$ $(i=1,2,3)$. 
We study here the mass spectra of these fields in order. 

\ben
\item Up-type quarks \\
As seen from the last term in Eq. (\ref{eqn:WYeff}) which contains 
$m_{ij} x^{\mu_{ij}} Q_i U^c_j H_{u0}$, 
the mass matrix for up-type quarks becomes 
\beq
  {\cal M}_{ij} v_u = m_{ij} \, x^{\mu_{ij}} v_u\,, 
\label{eqn:Mupq}
\eeq
where $v_u = \langle H_{u0} \rangle$ and $m_{ij} = \O(1)$. 
The exponents $\mu_{ij}$ are given in Eq. (\ref{eqn:power}). 
The eigenvalues of the mass matrix are 
\beq
  \O(x^{49} \, v_u), \qquad \O(x^{33} \, v_u), \qquad \O(x^3 \, v_u),
\eeq
which correspond to $m_u$, $m_c$ and $m_t$, respectively. 
Since $v_u = \O(10^2 \, {\rm GeV})$, 
all of up-type quarks remain massless in the intermediate 
energy region. 

\vspace{5mm}

\item Down-type colored fields \\
The effective superpotential of down-type colored fields 
with odd R-parity takes the form 
\beq
  W_D^{(eff)} = \sum_{i,j=1}^3 \left(
        z_{ij} \, x^{\zeta_{ij}} \, S_0 g_i g^c_j   
     +  m_{ij} \, x^{\mu_{ij}} \, N^c_0 g_i D^c_j 
     +  m_{ij} \, x^{\mu_{ij}} \, H_{d0} Q_i D^c_j \right). 
\eeq
When $S_0$, $N^c_0$ and $H_{d0}$ develop nonvanishing VEVs, 
the mass matrix is derived as 
\beq
\begin{array}{r@{}l} 
   \vphantom{\bigg(}   &  \begin{array}{ccc} 
          \quad \ \,  g^c   &  \quad \ D^c  &  
        \end{array}  \\ 
\widehat{{\cal M}}_{d} = 
   \begin{array}{l} 
        g  \\  D  \\   \end{array} 
     & 
\left( 
  \begin{array}{cc} 
     x^{0.5} {\cal Z}   &   y^{0.5} {\cal M}   \\
             0          &   \rho_d {\cal M}    \\
  \end{array} 
\right)
\end{array} 
\label{eqn:MD}
\eeq
in $M_1$ units. 
The submatrices ${\cal Z}$ and ${\cal M}$ 
are given by 
\beq
  {\cal Z}_{ij} = z_{ij} \, x^{\zeta_{ij}}
\eeq
and Eq. (\ref{eqn:Mupq}), respectively 
and $\rho_d = v_d/M_1$ with $v_d = \langle H_{d0} \rangle$. 
The mass matrix $\widehat{{\cal M}}_{d}$ yields mixings 
between $g^c$ and $D^c$ and has six eigenvalues. 
Three of them represent light modes. 
Solving the eigenvalue problem of this matrix, 
we obtain their masses\cite{Fuzz,Matsu1,CKM} 
\beq
  \O(x^{49} \, v_d), \qquad \O(x^{41} \, v_d), 
                         \qquad \O(x^{21} \, v_d),
\eeq
which correspond to $m_d$, $m_s$ and $m_b$, respectively. 
The remaining three represent heavy modes with their masses 
\beq
   \O(x^{13.125} \, M_1), \qquad \O(x^{31.5} \, M_1), \qquad 
   \O(x^{49.125} \, M_1).
\eeq

\vspace{5mm}

\item Charged leptons and extra charged Higgs fields with odd R-parity \\
For these fields the effective superpotential is written as 
\beq
  W_E^{(eff)} = \sum_{i,j=1}^3 \left(
           h_{ij} \, x^{\eta_{ij}} \, S_0 H_{di} H_{uj} 
         + m_{ij} \, x^{\mu_{ij}} \, N_0^c L_i H_{uj} 
         + m_{ij} \, x^{\mu_{ij}} \, H_{d0} L_i E^c_j \right). 
\label{eqn:WE}
\eeq
The mass matrix is similar to that for the down-type colored fields. 
Concretely, we have 
\beq
\begin{array}{r@{}l} 
   \vphantom{\bigg(}   &  \begin{array}{ccc} 
          \quad \  H_{u}   &  \quad \  E^c  &  
        \end{array}  \\ 
\widehat{{\cal M}}_{l} = 
   \begin{array}{l} 
        H_{d}  \\  L  \\   \end{array} 
     & 
\left( 
  \begin{array}{cc} 
     x^{0.5} {\cal H}    &        0            \\
     y^{0.5} {\cal M}    &  \rho_d {\cal M}    \\
  \end{array} 
\right)
\end{array} 
\label{eqn:Li}
\eeq
in $M_1$ units, where the submatrix ${\cal H}$ is given by 
\beq
  {\cal H}_{ij} = h_{ij} \, x^{\eta_{ij}}.
\eeq
The mass matrix $\widehat{{\cal M}}_{l}$ yields $L$-$H_d$ mixings. 
Among six eigenvalues of $\widehat{{\cal M}}_{l}$, 
three represent light modes corresponding to charged leptons. 
Their masses are given by\cite{Fuzz,Matsu1,MNS} 
\beq
  \O(x^{49} \, v_d), \qquad \O(x^{35} \, v_d), 
                         \qquad \O(x^{17} \, v_d),
\eeq
which represent $m_e$, $m_{\mu}$ and $m_{\tau}$, respectively. 
The remaining three represent heavy modes with their masses 
\beq
   \O(x^{2.5} \, M_1), \qquad \O(x^{29.125} \, M_1), \qquad 
   \O(x^{44.5} \, M_1).
\eeq

\vspace{5mm}

\item Neutral fields with odd R-parity \\
The neutral sector contains five types of matter fields, 
$H_u^0$, $H_d^0$, $L^0$, $N^c$ and $S$ in our model. 
Then we have the $15 \times 15$ mass matrix\cite{Fuzz,Matsu1,Matsu2,Matsu3,MNS} 
\begin{equation} 
\begin{array}{r@{}l} 
   \vphantom{\bigg(}   &  \begin{array}{cccccc} 
          \quad \, H_u^0   & \quad \  H_d^0  &  \qquad  L^0  
                          &  \quad \ \,  N^c   &  \quad \  S  &
        \end{array}  \\ 
\widehat{{\cal M}}_{NS} = 
   \begin{array}{l} 
        H_u^0  \\  H_d^0  \\  L^0  \\  N^c  \\  S  \\
   \end{array} 
     & 
\left( 
  \begin{array}{ccccc} 
       0     &  x^{0.5} {\cal H}   &    y^{0.5} {\cal M}^T     
                       &      0     &  \rho _d {\cal M}^T  \\
     x^{0.5} {\cal H}    &     0      &      0      
                       &      0     &  \rho _u {\cal M}^T  \\
     y^{0.5} {\cal M}    &     0     &      0      
                       &  \rho _u {\cal M} &       0       \\
       0     &     0     & \rho _u {\cal M}^T 
                       &      {\cal N}     &       0       \\
   \rho _d {\cal M} & \rho _u {\cal M} &      0      
                       &      0     &       {\cal S}       \\
  \end{array} 
\right) 
\end{array} 
\label{eqn:Mhn}
\end{equation}
in $M_S$ units, where $\rho _u = v_u/M_1$. 
In this mass matrix, the $3 \times 3$ submatrix ${\cal N}$ plays 
the role of the R-handed Majorana mass matrix in the seesaw mechanism. 
Among fifteen eigenvalues of $\widehat{{\cal M}}_{NS}$ 
three represent tiny masses of neutrinos,\cite{Fuzz,MNS} 
which are given by 
\beq
   \frac{v_u^2}{M_1 \, x^{37.25}} \times 
       (\O(x^{12}), \quad \O(x^4), \quad \O(1) \, ).
\eeq
Six of them are degenerate with the above-mentioned three heavy modes 
coming from extra charged Higgs fields. 
The remaining six eigenvalues are approximately given by 
the submatrices ${\cal N}$ and ${\cal S}$. 
${\cal N}$ is induced from the term 
\beq
  \frac{n_{ij}}{M_1} \, x^{\nu_{ij}} \, (N_i^c\bar{N^c})(N_j^c\bar{N^c})
\eeq
and of the form 
\beq
  {\cal N}_{ij} = n_{ij} \, x^{\nu_{ij}} \, y 
\eeq
in $M_1$ units, 
where the flavor symmetry leads to the relation 
$\nu_{ij} = \eta_{ij} + 49$. 
The three eigenvalues of ${\cal N}$ become 
\beq
   \O(x^{123.25} \, M_1), \qquad \O(x^{103.25} \, M_1), \qquad 
       \O(x^{71.25} \, M_1),
\eeq
which represent the masses of the R-handed Majorana neutrinos. 
The largest eigenvalue $\O(x^{71.25} \, M_1) \simeq 10^{10.8} \, $GeV 
is nearly equal to the geometrical average of $M_S$ and $M_Z$. 
The submatrix ${\cal S}$ induced from 
\beq
  \frac{s_{ij}}{M_1} \, x^{\sigma_{ij}} \, (S_i\bar{S})(S_j\bar{S})
\eeq
is given by 
\beq
  {\cal S}_{ij} = s_{ij} \, x^{\sigma_{ij} + 1} 
\eeq
in $M_1$ units, 
where we have the relation $\sigma_{ij} = \zeta_{ij} - 11$. 
The eigenvalues of ${\cal S}$ become 
\beq
   \O(x^{47} \, M_1), \qquad \O(x^{35} \, M_1), \qquad 
       \O(x^7 \, M_1).
\eeq

\een

\vspace{5mm}

\begin{table}[t]
\caption{The multiplicities and the spectra of the R-parity odd superfields}
\label{table:4}
\begin{center}
\renewcommand{\arraystretch}{1.1}
\begin{tabular}{|ccc|ccccccccc|} \hline \hline 
 & $\times M_1$ &   &   $Q$ & $U^c$ & $g$ & 
           $(D^c,g^c)$ & $E^c$ & $H_u$ & $(L,H_d)$ & $N^c$ & $S$  \\ 
\hline
  $1$          & $\sim$ & $x^{2.5}$     & 3 & 3 & 3 & 6 & 3 & 3 & 6 & 3 & 3 \\
  $x^{2.5}$    & $\sim$ & $x^7$         & 3 & 3 & 3 & 6 & 3 & 2 & 5 & 3 & 3 \\
  $x^7$        & $\sim$ & $x^{13.125}$  & 3 & 3 & 3 & 6 & 3 & 2 & 5 & 3 & 2 \\
  $x^{13.125}$ & $\sim$ & $x^{29.125}$  & 3 & 3 & 2 & 5 & 3 & 2 & 5 & 3 & 2 \\
  $x^{29.125}$ & $\sim$ & $x^{31.5}$    & 3 & 3 & 2 & 5 & 3 & 1 & 4 & 3 & 2 \\
  $x^{31.5}$   & $\sim$ & $x^{35}$      & 3 & 3 & 1 & 4 & 3 & 1 & 4 & 3 & 2 \\
  $x^{35}$     & $\sim$ & $x^{44.5}$    & 3 & 3 & 1 & 4 & 3 & 1 & 4 & 3 & 1 \\
  $x^{44.5}$   & $\sim$ & $x^{47}$      & 3 & 3 & 1 & 4 & 3 & 0 & 3 & 3 & 1 \\
  $x^{47}$     & $\sim$ & $x^{49.125}$  & 3 & 3 & 1 & 4 & 3 & 0 & 3 & 3 & 0 \\
  $x^{49.125}$ & $\sim$ & $x^{71.25}$   & 3 & 3 & 0 & 3 & 3 & 0 & 3 & 3 & 0 \\
  $x^{71.25}$  & $\sim$ & $x^{103.25}$  & 3 & 3 & 0 & 3 & 3 & 0 & 3 & 2 & 0 \\
  $x^{103.25}$ & $\sim$ & $x^{123.25}$  & 3 & 3 & 0 & 3 & 3 & 0 & 3 & 1 & 0 \\
  $x^{123.25}$ & $\sim$ &               & 3 & 3 & 0 & 3 & 3 & 0 & 3 & 0 & 0 \\
\hline
\end{tabular}
\end{center}
\vspace{3mm}
\end{table}

In Table 4 we summarize the multiplicities of the R-parity odd superfields. 
It is worthy to note that in the intermediate energy region we have 
the hierarchical spectra of heavy particles including the R-handed 
Majorana neutrinos. 

\vspace{10mm}

\section{RG equations and numerical study}
Now we study the two-loop RG evolutions of the gauge couplings 
in our model.
The evolution equations up to the two-loop order 
for $\alpha_i = g_i^2/4\pi$ are generally given by\cite{Falck} 
\beq
  \frac{d\alpha_i}{dt} = 
    \frac{1}{2\pi} \left[ -b_i + \frac{1}{4\pi}
       \left( \sum_j b_{ij} \alpha_j - a_i \right) 
          \right] \alpha_i^2,
\eeq
where $t=\ln(Q/Q_0)$ with $Q_0 = M_S$. 
The subscripts $i$ and $j$ specify the gauge group. 
The coefficients $b_i$, $b_{ij}$ and $a_i$ are determined 
depending on the particle contents and their spectra. 
The $a_i$ terms represent the contributions of 
Yukawa interactions.

In our model, the values of the gauge couplings 
$g_6$ of $SU(6)$ and $g_{2R}$ of $SU(2)_R$ are introduced 
at the string scale $M_S$ as an initial condition. 
As mentioned above, the gauge group $SU(6) \times SU(2)_R$ 
is broken down to $SU(4)_{PS} \times SU(2)_L \times SU(2)_R$ 
at the energy scale $|\langle \phi_0 \rangle| = x^{0.5} \, M_1$. 
The subsequent breaking of the gauge group into 
$SU(3)_c\times SU(2)_L\times U(1)_Y$ occurs at the energy scale 
$|\langle \psi_0 \rangle| = y^{0.5} \, M_1 \simeq x^{10.125} \, M_1$. 
The supersymmetry is supposed to be broken at the scale 
$\tilde{m}_{\phi} = 10^3 \, {\rm GeV} = x^{151} \, M_1$.
From the particle spectra given in Tables 3 and 4, 
we can determine the coefficients $b_i$ and $b_{ij}$ 
in various energy regions ranging from $M_S$ to $M_Z$.

In the first region between $M_S \ (t = 0)$ and 
$x^{0.5} \, M_1 \ (t = -1.21)$, 
where the gauge symmetry is $SU(6) \times SU(2)_R$, 
we have
\beq
    -b_i = \left(
       \begin{array}{r}
         -3 \\
          9 
       \end{array}
       \right),      \quad \
    b_{ij} =  \left(
       \begin{array}{cc}
         209  &  15  \\
         175  &  81 
       \end{array}
       \right),      \quad \
    a_i = \left(
       \begin{array}{c}
          8 \\
          0
       \end{array}
       \right)y_0
\eeq
with $y_0 = \vert z_0 \vert^2 /4\pi$. 
As seen from Eqs. (\ref{eqn:WY}) and (\ref{eqn:power}), 
in this region only $z_0 (\phi_0)^3$ term contributes to 
the RG equations for gauge couplings at two-loop level. 
The subscripts $i,j = 6$ and $2R$ denote $SU(6)$ and $SU(2)_R$, 
respectively. 
The one-loop RG equation for the Yukawa coupling $y_0$ 
is given by
\beq
  \frac{dy_0}{dt} = \frac{1}{2\pi} 
        \left( -28 \, \alpha_6 + 9 \, y_0 \right) y_0.
\eeq
As will be discussed later, the evolution of the gauge couplings 
are rather insensitive to the Yukawa couplings in the whole region. 
Then, in the present analysis it is sufficient for us 
to take the one-loop RG equations for the Yukawa couplings 
into account.

In the second region between $x^{0.5} \, M_1 \ (t = -1.21)$ and 
$x^{10.125} \, M_1 \ (t = -3.37)$, 
the gauge group is $SU(4)_{PS} \times SU(2)_L \times SU(2)_R$. 
In this region we obtain 
\bea
    -b_i &=& \left(
       \begin{array}{c}
             0     \\
            n_H    \\
          4 + n_H 
       \end{array}
       \right),      \quad \
    b_{ij} \ = \ \left(
       \begin{array}{ccc}
         100  &     9       &     15      \\
          45  &  18 + 7n_H  &    3n_H     \\
          75  &    3n_H     &  46 + 7n_H 
       \end{array}
       \right),     \nonumber \\
    a_i &=& \left(
       \begin{array}{c}
         4 \, (y^{(Z)} + 2y^{(1)} + 2y^{(2)} + 4y^{(3)})            \\
         4 \, (y^{(H)} + 4y^{(1)} + 4y^{(2)} +  y^{(4)})            \\
         4 \, (y^{(H)} + 4y^{(1)} + 4y^{(2)} + 3y^{(3)} + y^{(4)})
       \end{array}
       \right), 
\label{eqn:Secbij}
\eea
where $i,j = 4,2L$ and $2R$ means $SU(4)_{PS}$, $SU(2)_L$ and $SU(2)_R$, 
respectively. 
The notation $n_H$ in Eq. (\ref{eqn:Secbij}) represents 
the multiplicity of doublet Higgs fields and is given by 
\beq
  n_H = \left\{
        \begin{array}{ll}
          5 \qquad x^{0.5} > Q/M_1 \geq x^{2.5},  \\
          4 \qquad x^{2.5} > Q/M_1 \geq x^{10.125}.
        \end{array}
       \right.
\eeq
At $x^{0.5} \, M_1$ we use the continuity condition 
\beq
 \alpha_6 = \alpha_4 = \alpha_{2L}.
\eeq
The dominant contributions of the effective Yukawa interactions 
are of the forms 
\bea
   y^{(Z)} & = & \frac{1}{4\pi} \vert f_{33}^{(Z)} \vert ^2, \qquad 
   y^{(1)} \ = \ \frac{1}{4\pi} \vert M_{33}^{(1)} \vert ^2, \qquad 
   y^{(3)} \ = \ \frac{1}{4\pi} \vert M_{33}^{(3)} \vert ^2,  \nonumber \\
   y^{(H)} & = & \frac{1}{4\pi} \vert f_{33}^{(H)} \vert ^2 \, 
          \theta\left(\frac{Q}{M_1} - x^{2.5}\right), \quad  
   y^{(2)} \ = \ \frac{1}{4\pi} \vert M_{33}^{(2)} \vert ^2 \, 
          \theta\left(\frac{Q}{M_1} - x^{2.5}\right),   \nonumber \\
   y^{(4)} & = & \frac{1}{4\pi} \vert M_{33}^{(4)} \vert ^2 \, 
          \theta\left(\frac{Q}{M_1} - x^{2.5}\right).
\eea
The definition of the effective Yukawa couplings $f_{33}^{(Z,H)}$ 
and $M_{33}^{(1 \sim 4)}$ are presented in the Appendix. 
The RG equations for the Yukawa couplings are also given 
in the Appendix.

In the third region ranging from $x^{10.125} \, M_1 \ (t = -3.37)$ to 
$x^{151} \, M_1(= \tilde{m}_{\phi} = 10^3 \, {\rm GeV})$ $(t = -35.00)$, 
where the gauge group coincides with 
$G_{SM} = SU(3)_C \times SU(2)_L \times U(1)_Y$, 
still being supersymmetric, 
we obtain
\bea
    -b_i &=& \left(
       \begin{array}{c}
          -3 + N_g      \\
               n_H      \\
           6 + \frac{2}{5}N_g + \frac{3}{5}n_H 
       \end{array}
       \right),               \nonumber \\
    b_{ij} &=&  \left(
       \begin{array}{ccc}
           14   &    9   &  11/5  \\
           24   &   18   &   6/5  \\
          88/5  &  18/5  &  38/5 
       \end{array}
       \right) 
       +  \left(
       \begin{array}{ccc}
          34/3   &  0  &  4/15  \\
            0    &  0  &    0   \\
          32/15  &  0  &  8/75 
       \end{array}
       \right)N_g             \nonumber \\
       & & \phantom{MMM} +  \left(
       \begin{array}{ccc}
          0  &   0   &   0   \\
          0  &   7   &  3/5  \\
          0  &  9/5  &  9/25 
       \end{array}
       \right)n_H,            \nonumber \\
    a_i &=& \left(
       \begin{array}{c}
           4  \\
           6  \\
         26/5 
       \end{array}
       \right)y^{(1U)} +  \left(
       \begin{array}{c}
           4  \\
           6  \\
         14/5 
       \end{array}
       \right)y^{(1D)}        \nonumber \\
       & & \phantom{MMM} +  \left(
       \begin{array}{c}
          1   \\
          0   \\
         2/5 
       \end{array}
       \right)\left(y^{(3N)} + 2y^{(3D)} +2y'^{(Z)}\right), 
\eea
where $i,j = 3,2L$ and $1$ means $SU(3)_c$, $SU(2)_L$ and $U(1)_Y$, 
respectively. 
$n_H$ and $N_g$, the latter being the multiplicity of the extra 
down-type colored fields, are shown in Table 5. 
The Yukawa terms are expressed as 
\bea
   y^{(1U)} &=& \frac{1}{4\pi}\vert M_{33}^{(1U)}\vert ^2,\qquad 
   y^{(1D)} =   \frac{1}{4\pi}\vert M_{33}^{(1D)}\vert ^2 \, 
          \theta\left(\frac{Q}{M_1} - x^{13.125}\right), \nonumber \\
   y^{(3N)} &=& \frac{1}{4\pi}\vert M_{33}^{(3N)}\vert ^2 \, 
          \theta\left(\frac{Q}{M_1} - x^{13.125}\right), \quad
   y^{(3D)} =   \frac{1}{4\pi}\vert M_{33}^{(3D)}\vert ^2 \, 
          \theta\left(\frac{Q}{M_1} - x^{13.125}\right), \nonumber \\ 
   y^{'(Z)} &=& \frac{1}{4\pi}\vert f_{33}^{'(Z)}\vert ^2 \, 
          \theta\left(\frac{Q}{M_1} - x^{13.125}\right).
\eea
The continuity conditions at the scale $x^{10.125} \, M_1$ are given by 
\bea
  \alpha_4      &=& \alpha_3,                             \\
  \alpha_1^{-1} &=& \frac{2}{5} \, \alpha_4^{-1} 
                      + \frac{3}{5} \, \alpha_{2R}^{-1},  \\
  y^{(1)}       &=& y^{(1U)} = y^{(1D)},                  \\
  y^{(3)}       &=& y^{(3N)} = y^{(3D)},                  \\
  f_{33}^{'(Z)} &=& x^{4.375}f_{33}^{(Z)} 
                   + \frac{1}{\sqrt{2}} \, M_{33}^{(3)}.
\eea
The last condition is due to the mixing of $g^c$ and 
$D^c$ as seen in Eq. (\ref{eqn:MD}). 
In the Appendix we list the definitions of $f_{33}^{'(Z)}$ 
and $M_{33}^{(1U \sim 3D)}$ and the RG equations for 
the Yukawa couplings.

\begin{table}[t]
\caption{The values of $n_H$ and $N_g$ in the region between 
$x^{10.125} \, M_1$ and $x^{151} \, M_1$. 
$n_H$ and $N_g$ are the multiplicities of doublet Higgses and 
extra down-type colored fields, respectively. 
Here, the parameter $k$ is taken as $31.5<k<44.5$. 
In the numerical calculation, we use the adjusted value $k = 42.5$.}
\label{table:5}
\begin{center}
\renewcommand{\arraystretch}{1.1}
\begin{tabular}{|ccc|cc|} \hline \hline
  & $\times M_1$  &  & \phantom{M} $n_H$ \phantom{M} & 
               \phantom{M} $N_g$ \phantom{M} \\  
\hline 
$x^{10.125}$  & $\sim$ & $x^{13.125}$   &    4     &     5    \\
$x^{13.125}$  & $\sim$ & $x^{29.125}$   &    4     &     4    \\
$x^{29.125}$  & $\sim$ & $x^{31.5}$     &    3     &     4    \\
$x^{31.5}$    & $\sim$ & $x^k$          &    3     &     3    \\
$x^k$         & $\sim$ & $x^{44.5}$     &    3     &     2    \\
$x^{44.5}$    & $\sim$ & $x^{49.125}$   &    2     &     2    \\
$x^{49.125}$  & $\sim$ & $x^{125}$      &    2     &     1    \\
$x^{125}$     & $\sim$ & $x^{141.75}$   &    1     &     1    \\
$x^{141.75}$  & $\sim$ & $x^{151}$      &    1     &     0    \\
\hline
\end{tabular}
\end{center}
\vspace{3mm}
\end{table}

The final region with the gauge symmetry $G_{SM}$ is between 
$x^{151} \, M_1 (t = -35.00)$ 
and $x^{161.4} \, M_1( = M_Z)$ $(t = -37.34)$,
where the supersymmetry is broken. 
In this region, we have
\bea
    -b_i &=& \left(
       \begin{array}{c}
               -7               \\
          -\frac{10}{3} + n_H   \\
            4 + \frac{3}{5}n_H 
       \end{array}
       \right),    \nonumber \\
    b_{ij} &=&  \left(
       \begin{array}{ccc}
          -26   &   9/2  &  11/10  \\
           12   &  11/3  &   3/5   \\
          44/5  &   9/5  &  19/5 
       \end{array}
       \right) 
       +  \left(
       \begin{array}{ccc}
          0  &    0    &   0    \\
          0  &  25/2   &  9/10  \\
          0  &  27/10  &  27/50 
       \end{array}
       \right)n_H,  \nonumber \\
    a_i &=& \left(
       \begin{array}{c}
           2    \\
          3/2   \\
         17/10 
       \end{array}
       \right)y^{(1U)},
\eea
where
\beq
 n_H = \left\{
        \begin{array}{ll}
          1 \qquad x^{151} > Q/M_1 \geq x^{158},  \\
          0 \qquad x^{158} > Q/M_1 \geq x^{161.4}. 
        \end{array}
       \right.
\eeq
The evolution equation for the Yukawa coupling is given by
\beq
  \frac{dy^{(1U)}}{dt}  =  \frac{1}{2\pi} \left[ 
      -8\alpha_3 - \frac{9}{4}\alpha_{2L} - \frac{17}{20}\alpha_1 
                + \frac{9}{2}y^{(1U)} \right] y^{(1U)}.
\eeq

We are now in a position to solve the RG equations numerically. 
By adjusting the relevant parameters, 
we can obtain some consistent solutions of the gauge unification. 
A typical solution is shown in Fig. 1, 
where the initial values of gauge coupling constants 
at the string scale $M_S$ are taken as 
\beq
   \alpha_6(t=0) = 0.057, \qquad 
   \alpha_{2R}(t=0) = 0.083 
\eeq
and that of the Yukawa coupling as $y_0(t=0) = 0.6$. 
In this solution the values of $z_{33}$, $h_{33}$ and $m_{33}$ 
in Eq. (\ref{eqn:WYeff}) are taken as 1.0, 0.3 and 2.0, respectively, 
and the parameter $k$ is adjusted as $k = 42.5$. 
Under this choice of the parameter values, 
we obtain the running coupling constants at $M_Z$ as
\bea
  \alpha_3^{-1}(M_Z)    &=&  8.44, \\
  \alpha_{2L}^{-1}(M_Z) &=& 29.79, \\
  \alpha_1^{-1}(M_Z)    &=& 59.48. 
\eea
These results are in good agreement with 
the experimental values.\cite{PDG} 
Numerically, the results are insensitive to the values of 
$z_{33}$, $h_{33}$ and $m_{33}$ but appreciably affected by 
the value of $k$. 
For instance, if we take $k=37.5$ instead of $k=42.5$, 
the coupling constants are changed as 
\bea
  \alpha_3^{-1}(M_Z)    &=&  8.26, \\
  \alpha_{2L}^{-1}(M_Z) &=& 29.79, \\
  \alpha_1^{-1}(M_Z)    &=& 59.41. 
\eea
In our model the spectra of the anti-generation fields 
$\bar{g}$ and $\bar{g^c}$ play an important role in adjusting 
the gauge couplings. 
Since the fields $\bar{g}$ and $\bar{g^c}$ are colored but 
$SU(2)_L$-singlets, 
$\alpha_3(M_Z)$ and $\alpha_1(M_Z)$ increase with decreasing $k$ 
but $\alpha_{2L}(M_Z)$ is almost independent of $k$. 
We find a consistent solution of the gauge coupling unification, 
which represents the connection between the string scale physics 
and the electroweak scale physics.

\begin{figure}[t] 
 \centerline{\includegraphics[width=11 cm,height=8 cm]
{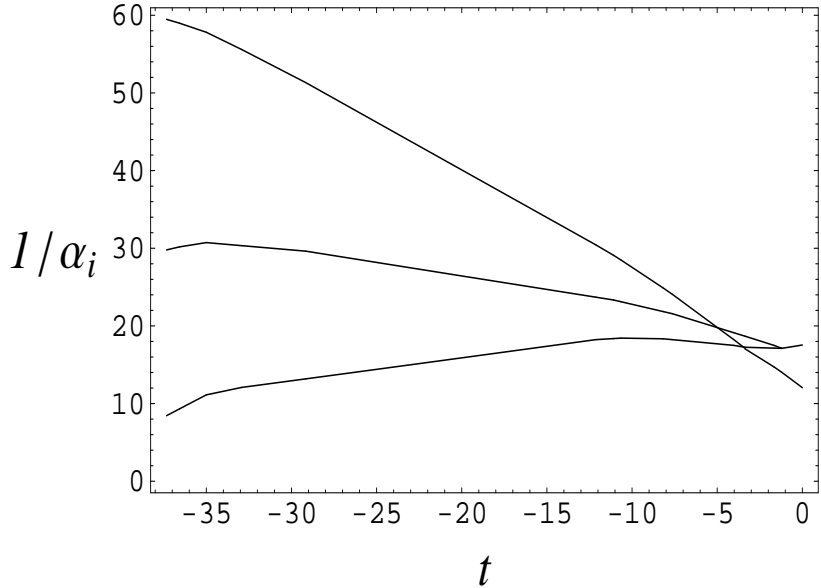}}
\caption{
Running inverse gauge couplings up to two-loop effects 
in our model. 
Vertical and horizontal axes show $\a_i^{-1}$ and 
$t=\ln(Q/M_S)$, respectively. 
The upper, middle and bottom lines in the left hand region 
correspond to $\a_1^{-1}$, $\a_2^{-1}$ and $\a_3^{-1}$, respectively. 
The initial values of Yukawa couplings are taken as 
$z_{33}=1.0$, $h_{33}=0.3$ and $m_{33}=2.0$.
} 
\label{fig:1}
\vspace{3mm}
\end{figure} 

\vspace{10mm}

\section{Summary and discussion}
Characteristic patterns in fermion masses and mixings 
at low energies strongly suggest the existence of 
a profound type of the flavor symmetry including the R-parity. 
It is plausible that the flavor symmetry also controls 
the mass spectra of heavy particles which appear 
on the way between the string scale and the electroweak scale. 
In order to explore the path for connecting the string scale physics 
with the low-energy physics, it is necessary for us to study 
these spectra of heavy particles. 
For this purpose we examined particle spectra in the context 
of the $SU(6) \times SU(2)_R$ string-inspired model 
with the flavor symmetry 
${\bf Z}_{19} \times {\bf Z}_{18} \times \tilde{D}_4$. 
In our model the gauge symmetry is spontaneously broken 
in two steps as 
\[
   SU(6) \times SU(2)_R 
     \buildrel \langle \phi_0 \rangle \over \longrightarrow 
             SU(4)_{\rm PS} \times SU(2)_L \times SU(2)_R  
             \buildrel \langle \psi_0 \rangle \over \longrightarrow 
     G_{\rm SM}, 
\]
where 
$| \langle \phi_0 \rangle | \sim 4.5 \times 10^{17} \, {\rm GeV}$ and 
$| \langle \psi_0 \rangle | \sim 5.2 \times 10^{16} \, {\rm GeV}$. 
It was shown that there appear characteristic patterns of spectra 
also in the intermediate energy region. 
Afterward, we gave the two-loop RG equations of the gauge 
couplings in the intermediate region, in which each heavy particle 
decouples by its own stage. 
The RG runnings of the gauge couplings were studied 
up to two-loop order and we have obtained consistent gauge 
couplings at $M_Z$ with the experimental values by taking 
the reasonable values for the available parameters.  
We explored solutions in which $SU(3)_c$ and $SU(2)_L$ 
gauge couplings meet at $\O(5 \times 10^{17} \, {\rm GeV})$ 
and found a solution of the gauge coupling unification 
by adjusting the spectra of the anti-generation matter fields. 
The solution represents the connection between 
the string scale physics and the electroweak scale physics. 
It should be emphasized that in our model 
the unification scale is around the string scale 
($\sim 10^{18} \, $GeV) but not the so-called GUT scale 
($\sim 2 \times 10^{16} \, $GeV).

In the present analysis the gauge couplings at the string 
scale $M_S = 1.5 \times 10^{18} \, $GeV are numerically taken as 
\beq
   \alpha_6(t=0) = 0.057, \qquad \alpha_{2R}(t=0) = 0.083. 
\eeq
This means that the gauge couplings of $SU(6)$ and $SU(2)_R$ 
are not unified in the four-dimensional effective theory. 
In the framework of the higher-dimensional underlying theory, 
however, the gauge unification of $SU(6)$ and $SU(2)_R$ 
is expected to be realized non-perturbatively. 
There is a possibility that the $SU(6)$ and $SU(2)_R$ gauge groups 
live in distinct world-volumes of different D-branes. 
It has been pointed out that if the $SU(6)$ gauge group lives in 
the world-volume of 9-branes and the $SU(2)_R$ in the world-volume 
of 5-branes, or {\it vice versa}, unification solutions can be 
found in the vicinity of the self-dual point 
in the moduli space.\cite{SDP}

In the present study we parameterized the magnitude of 
the effective Yukawa interactions ${\bar \phi}^3$ 
to adjust the spectra of the anti-generation fields. 
In order to determine the spectra of the anti-generation fields, 
we need to know the flavor charges of the K\"ahler class 
moduli fields. 
To this end, it is necessary to elucidate the origin of 
the flavor symmetry linked to the structure of the compact space. 
Interestingly, in the case of the four-generation model 
we were able to obtain the flavor symmetry and to calculate 
the flavor charges of the matter superfields, 
explicitly.\cite{4gen} 
In this case the compact space is a quintic hypersurface 
in $CP^4$. 
String compactification on this hypersurface corresponds to 
a deformation of the $3^5$ Gepner model.\cite{Gepner} 
In addition, it is considered that the flavor symmetry has 
its origin not only in the symmetric structure of the compact space 
but also in the non-commutativity in the compact space.\cite{Fuzz} 
In the context of M-theory on $G_2$ holonomy spaces, 
the gauge symmetry at the string scale is related to 
the singularity structure of the compact space\cite{Acharya} 
and constrained by the cohomology condition on the brane 
configuration.\cite{Ovrut} 
In this context the flavor symmetry might be also constrained 
depending on the singularity structure of the compact space. 
The flavor symmetry is expected to provide an important clue 
to the study of the path for connecting the string scale physics 
with the low-energy physics.

\vspace{10mm}

\section*{Appendix}
\setcounter{equation}{0}
\renewcommand{\theequation}{A.\arabic{equation}}
By using the effective superpotential Eq. (\ref{eqn:WYeff}), 
we can extract the relevant Yukawa terms which give 
the dominant contributions to the RG equations.

Due to the symmetry breaking of $SU(6) \times SU(2)_R$ 
down to $SU(4)_{PS} \times SU(2)_L \times SU(2)_R$, 
the matter superfields $\phi({\bf 15, \ 1})$ and 
$\psi({\bf 6^*, \ 2})$ are decomposed as 
\bea
  \phi({\bf 15, \ 1}) & \longrightarrow & \phi({\bf 4, \ 2, \ 1}) 
              + \phi({\bf 6, \ 1, \ 1}) + \phi({\bf 1, \ 1, \ 1}), 
                                                      \nonumber \\
  \psi({\bf 6^*, \ 2}) & \longrightarrow & \psi({\bf 4^*, \ 1, \ 2}) 
              + \phi({\bf 1, \ 2, \ 2}), 
\eea
respectively in the region between $x^{0.5} \, M_1$ and $x^{10.125} \, M_1$. 
Therefore, the effective superpotential is expressed as 
\bea
   W_Y^{(eff)} & = & \frac{1}{2} f^{(\bar{Z})} 
                      \tilde{S} \, \bar{\phi(6,1,1)}^2 
          + \frac{1}{2} f^{(H_0)} 
                      \tilde{S} \, \psi(1,2,2)_0^2     \nonumber \\
     {} & &  + \frac{1}{2} f_{ij}^{(Z)} \tilde{S} \, 
                      \phi(6,1,1)_i \, \phi(6,1,1)_j 
          + \frac{1}{2} f_{ij}^{(H)} \tilde{S} \, 
                      \psi(1,2,2)_i \, \psi(1,2,2)_j   \nonumber \\
     {} & &  + \ M_{ij}^{(1)} \psi(1,2,2)_0 \, 
                      \phi(4,2,1)_i \, \psi(4^*,1,2)_j \nonumber \\
     {} & &  + \ M_{ij}^{(2)} \psi(4^*,1,2)_0 \, 
                      \phi(4,2,1)_i \, \psi(1,2,2)_j   \nonumber \\
     {} & &  + \ M_{ij}^{(3)} \psi(4^*,1,2)_0 \, 
                      \phi(6,1,1)_i \, \psi(4^*,1,2)_j \nonumber \\
     {} & &  + \ M_{ij}^{(4)} \psi(1,2,2)_0 \, 
                      \phi(1,1,1)_i \, \psi(1,2,2)_j, 
\label{eqn:AWeff}
\eea
where the effective Yukawa couplings are of the forms 
\bea
   f^{(\bar{Z})} &=& \frac{(2\bar{\zeta} + 1)}{\sqrt{2}} \, 
                         \bar{z} \, x^{\bar{\zeta}}, \phantom{MMM} 
   f^{(H_0)}     \ = \ \frac{(2\eta_0 + 1)}{\sqrt{2}} \, 
                         h_0 \, x^{\eta_0},         \nonumber \\
   f_{ij}^{(Z)}  &=& \frac{(2\zeta_{ij}+ 1)}{\sqrt{2}} \, 
                         z_{ij} \, x^{\zeta_{ij}},   \qquad 
   f_{ij}^{(H)}  \ = \ \frac{(2\eta_{ij} + 1)}{\sqrt{2}} \, 
                         h_{ij} \, x^{\eta_{ij}} 
\eea
and
\beq
   M_{ij}^{(a)} = m_{ij}^{(a)} \, x^{\mu_{ij}}. \qquad (a=1,2,3,4) 
\eeq
As to the last equation, note that $M_{ij}^{(a)} (a = 1 \sim 4)$ 
evolves separately, but 
\beq
  M_{ij}^{(1)} = M_{ij}^{(2)} = M_{ij}^{(3)} = M_{ij}^{(4)} 
                      = m_{ij} \, x^{\mu_{ij}} 
\eeq
at scale $|\langle \phi_0 \rangle| = x^{0.5} \, M_1$. 
As seen in Eq. (\ref{eqn:power}), 
the dominant contributions in the effective potential 
come from the terms with small powers of $x$. 
It turns out that such terms are the effective Yukawa interactions 
with $f_{33}^{(Z)}$, $f_{33}^{(H)}$ and $M_{33}^{(1 \sim 4)}$ 
in Eq. (\ref{eqn:AWeff}). 
The RG equations for the Yukawa couplings are given by 
\bea
  \frac{dy^{(Z)}}{dt} & = & \frac{1}{2\pi}\left[ 
     -10\alpha_4 + 8y^{(Z)} + 2y^{(H)} +4y^{(3)} 
                               \right] y^{(Z)},     \nonumber \\
  \frac{dy^{(H)}}{dt} & = & \frac{1}{2\pi}\left[ 
     -3\alpha_{2L} - 3\alpha_{2R} + 6y^{(Z)} 
                  + 4y^{(H)} + 8y^{(2)} + 2y^{(4)} 
                               \right] y^{(H)},     \nonumber \\
  \frac{dy^{(1)}}{dt} & = & \frac{1}{2\pi}\left[ 
     -\frac{15}{2}\alpha_4 - 3\alpha_{2L} - 3\alpha_{2R} 
       + 8y^{(1)} + 2y^{(2)} + 3y^{(3)} + y^{(4)} 
                               \right] y^{(1)},     \nonumber \\
  \frac{dy^{(2)}}{dt} & = & \frac{1}{2\pi}\left[ 
     -\frac{15}{2}\alpha_4 - 3\alpha_{2L} - 3\alpha_{2R} 
       + y^{(H)} + 2y^{(1)} + 8y^{(2)} + 3y^{(3)} + y^{(4)} 
                               \right] y^{(2)},     \nonumber \\
  \frac{dy^{(3)}}{dt} & = & \frac{1}{2\pi}\left[ 
     -\frac{25}{2}\alpha_4 - 3\alpha_{2R} 
       + y^{(Z)} + 2y^{(1)} + 2y^{(2)} + 7y^{(3)} 
                               \right] y^{(3)},     \nonumber \\
  \frac{dy^{(4)}}{dt} & = & \frac{1}{2\pi}\left[ 
     -3\alpha_{2L} - 3\alpha_{2R} 
       + y^{(H)} + 4y^{(1)} + 4y^{(2)} + 6y^{(4)} 
                               \right] y^{(4)}.
\eea

In the region between $x^{10.125} \, M_1$ and $x^{151} \, M_1$, 
where the gauge group is 
$G_{SM} = SU(3)_c \times SU(2)_L \times U(1)_Y$, 
the matter superfields are further decomposed as 
\bea
  \phi({\bf 4, \ 2, \ 1}) & \longrightarrow & \phi({\bf 3, \ 2, \ 1/3}) 
                            + \phi({\bf 1, \ 2, \ -1}),    \nonumber \\
  \phi({\bf 6, \ 1, \ 1}) & \longrightarrow & \phi({\bf 3, \ 1, \ -2/3}) 
                           + \phi({\bf 3^*, \ 1, \ 2/3}),  \nonumber \\
  \phi({\bf 1, \ 1, \ 1}) & \longrightarrow & \phi({\bf 1, \ 1, \ 0}), 
                                                           \nonumber \\
  \psi({\bf 4^*, \ 1, \ 2}) & \longrightarrow & 
    \psi({\bf 3^*, \ 1, \ -4/3}) + \phi({\bf 3^*, \ 1, \ 2/3}) \nonumber \\
         {} & & \phantom{MMMMM} + \psi({\bf 1, \ 1, \ 2}) 
                          + \phi({\bf 1, \ 1, \ 0}),       \nonumber \\
  \psi({\bf 1, \ 2, \ 2}) & \longrightarrow & \psi({\bf 1, \ 2, \ 1}) 
                            + \phi({\bf 1, \ 2, \ -1}).
\eea
In this region we obtain five dominant Yukawa terms 
\bea
   M_{33}^{(1U)} \, 
      \tilde{H}_{u0} \, \tilde{Q_3} \, \tilde{U_3^c}, & \qquad & 
   M_{33}^{(1D)} \, 
      \tilde{H}_{d0} \, \tilde{Q_3} \, \tilde{g_3^c},       \nonumber \\
   \frac{1}{\sqrt{2}}M_{33}^{(3N)} \, 
               \tilde{N^c} \, \tilde{g_3} \, \tilde{g_3^c}, & \qquad &
   M_{33}^{(3D)} \, 
      \tilde{D_0^c} \, \tilde{g_3} \, \tilde{N_3^c}         \nonumber 
\eea
and
\beq
  f_{33}^{'(Z)} \, \tilde{S} \, \tilde{g_3} \, \tilde{g_3^c} = 
          (x^{4.375}f_{33}^{(Z)}+ \frac{1}{\sqrt{2}}M_{33}^{(3)})
               \tilde{S} \, \tilde{g_3} \, \tilde{g_3^c}. 
\eeq
We use here the notation "$\sim$" for mass eigenstates. 
The last equation in the above comes from the $g^c$-$D^c$ mixing. 
The RG equations for these Yukawa couplings are given by 
\bea
  \frac{dy^{(1U)}}{dt} & = & \frac{1}{2\pi}\left[ 
     -\frac{16}{3}\alpha_3 - 3\alpha_{2L} - \frac{13}{15}\alpha_1 
        + 6y^{(1U)} + y^{(1D)} 
                             \right] y^{(1U)},           \nonumber \\
  \frac{dy^{(1D)}}{dt} & = & \frac{1}{2\pi}\left[ 
     -\frac{16}{3}\alpha_3 - 3\alpha_{2L} - \frac{7}{15}\alpha_1 
        + y^{(1U)} + 6y^{(1D)} + \frac{1}{2}y^{(3N)} + y^{'(Z)} 
                             \right] y^{(1D)},           \nonumber \\
  \frac{dy^{(3N)}}{dt} & = & \frac{1}{2\pi}\left[ 
     -\frac{16}{3}\alpha_3 - \frac{4}{15}\alpha_1 
        + 2y^{(1D)} +  5y^{(3N)} + y^{(3D)} + 2y^{'(Z)} 
                             \right] y^{(3N)},           \nonumber \\
  \frac{dy^{(3D)}}{dt} & = & \frac{1}{2\pi}\left[ 
     -\frac{16}{3}\alpha_3 - \frac{4}{15}\alpha_1 
        + \frac{1}{2}y^{(3N)} + 5y^{(3D)} + y^{'(Z)} 
                             \right] y^{(3D)},           \nonumber \\
  \frac{dy^{'(Z)}}{dt} & = & \frac{1}{2\pi}\left[ 
     -\frac{16}{3}\alpha_3 - \frac{4}{15}\alpha_1 
        + 2y^{(1D)} + y^{(3N)} + y^{(3D)} + 5y^{'(Z)} 
                             \right] y^{'(Z)}.
\eea

\vspace{10mm}

\section*{Acknowledgements}
Two of the authors (M. M. and T. M.) are supported in part by 
a Grant-in-Aid for Scientific Research from the Ministry of 
Education, Culture, Sports, Science and Technology, 
Japan (No.~12047226).



\end{document}